\documentclass[11pt]{article}
\usepackage[margin=1in]{geometry}
\usepackage{amsmath,amssymb,amsthm}
\usepackage{graphicx}
\usepackage{booktabs}
\usepackage[hidelinks]{hyperref}
\usepackage{mathtools}
\usepackage[section]{placeins}

\title{Minimum settling-time PI control of pure delay processes under a hard non-overshoot constraint: exact boundary-contact characterization and the role of the MID point}
\author{Senol Gulgonul\\ Ostim Technical University, Ankara, Turkey}
\date{}
\begin{document}
\maketitle

\begin{abstract}

We solve, exactly, the problem of minimum settling-time PI control of a pure delay process $K e^{-Ls}$ under the hard time-domain constraint of zero overshoot, $y(t)\le 1$ for all $t$. The closed loop is a neutral delay system whose step response is piecewise polynomial on the delay segments and carries geometrically decaying jump discontinuities at the segment boundaries $t=kL$. As a consequence, the constrained optimum is characterized by an equioscillation-type contact structure in which the active contacts sit at echo boundaries: kink maxima grazing the setpoint, jumps landing exactly on the settling-band edge, and boundary troughs anchored to it. The number of active contact equations equals the number of controller gains, rendering the optimum exactly computable for every band $\delta$. In a closed-form regime, $\delta\in[(3-2\sqrt{2})/4,\,(3-2\sqrt{2})/2]\approx[4.29\%,\,8.58\%]$, the optimal gains are independent of $\delta$, namely $KK_p=1-\sqrt{2}/2$ and $KK_iL=\sqrt{2}/2$ with $T_i=(\sqrt{2}-1)L$, and the optimal settling time is $T_s^*(\delta)=(4-\sqrt{2}-2\sqrt{\delta})L$. Outside this window the optimum is the root of an explicit two-equation polynomial system per regime, and $T_s^*(\delta)$ is a staircase with exact flats at integer multiples of $L$ produced by jump-landing pinning. As $\delta\to 0$ the optimal gains converge to $KK_p=e^{-2}$, $KK_iL=4e^{-2}$; this is exactly the generic multiplicity-induced-dominancy (GMID) point of the neutral quasipolynomial recently characterized by Boussaada, Mazanti and Niculescu. We show that the GMID step response itself satisfies the hard constraint and is the unique non-overshooting maximizer of the decay rate; yet, at every finite $\delta$ down to $10^{-6}$, the $\delta$-adapted contact optimum strictly beats the fixed GMID tuning, by about 40\% at $\delta=2\%$ and by a margin consistent with the $\Theta(\ln\ln(1/\delta))$ secular penalty of the triple root. The MID point is therefore the limit of the optimal gains without ever being the optimal tuning. Comparisons with the pole-zero-cancellation (Lambert-W) family of the companion paper and a numerical extension to first-order-plus-time-delay (FOTD) plants quantify the speed/robustness trade across $M_s\in[1.39,\,1.76]$.


\end{abstract}
\section{Introduction}

Proportional-integral (PI) and proportional-integral-derivative (PID) tuning for delay-dominated processes is classically driven by integral performance criteria such as the integral of squared error (ISE), integral of absolute error (IAE) and integral of time-weighted absolute error (ITAE) and their descendants \cite{r1}. A more recent line is spectral: placing or bounding the rightmost characteristic roots of the closed-loop quasipolynomial. This line has crystallized into the multiplicity-induced-dominancy (MID) program, in which a characteristic root of maximal multiplicity is, for broad classes of single-delay systems, necessarily dominant, so that forcing maximal multiplicity yields an analytically assigned decay rate together with a certified spectrum location \cite{r2,r3,r4}. In parallel, it was shown that for retarded quasipolynomials the maximum decay rate over a controller parameter region need not occur at a multiple root at all, and a $\sigma$-stability contraction methodology was given to locate the true points of best performance \cite{r5}.

This literature is entirely spectral. Two questions that matter directly to the practitioner are not addressed by it.

The first question concerns hard time-domain constraints. Industrial setpoint specifications are stated on the response itself, in the form "no overshoot, settle into $\pm\delta$ as fast as possible", not on the spectral abscissa. For finite $\delta$ the settling time of a delay system is not a function of the abscissa alone; it is governed by the early transient segments, which for delay systems have a specific echo structure. To our knowledge there is no exact treatment, for any delay plant, of the problem of minimizing $T_s(\delta)$ over $(K_p,K_i)$ subject to $y(t)\le 1$ for all $t\ge 0$.

The second question concerns the relation between this constrained problem and the spectral optimum. Since the step error of an exponentially stable delay system decays at every rate below its spectral abscissa $\alpha$ \cite{r6}, one expects $T_s(\delta)\approx \ln(1/\delta)/|\alpha|$ as $\delta\to 0$, so the limit of the constrained problem should connect to constrained abscissa optimization. Whether that limit is the MID point, whether the MID point even satisfies the non-overshoot constraint, and whether the MID tuning is ever the actual minimizer at finite $\delta$, are open. Notably, for neutral quasipolynomials, which is what PI control of a relative-degree-zero plant produces, the maximum-damping question is itself flagged as open in \cite{r3}, and the methodology of \cite{r5} formally assumes retarded type and does not apply.

This paper answers both questions for the pure delay plant. First, we give an exact boundary-contact characterization of the constrained optimum: the response carries geometrically decaying jumps at the echo boundaries, the active contacts of the optimal response sit at these boundaries, and their number equals the number of gains, so each tolerance regime reduces to an explicit small polynomial system (Theorem 1). Second, a closed-form regime exists in which the optimal gains are independent of the band and the optimal settling time is an elementary expression (Theorem 2). Third, as the band shrinks the optimal gains converge to the GMID point of the closed-loop quasipolynomial, recently characterized in \cite{r3}; we show that this spectrally defined point satisfies the hard constraint and uniquely maximizes the decay rate, yet the adapted contact optimum strictly beats the fixed GMID tuning at every finite band, so the MID point is the limit of the minimizers without ever being the minimizer. Fourth, a design ladder quantifies the speed and robustness trade against the cancellation family of the companion paper \cite{r7}, and a numerical FOTD extension shows that the structure persists, with a non-monotone settling time in the delay-to-lag ratio.

Relative to the cancellation-based analytical design of the companion paper, with structured gains $K_p=TK_i$ and a Lambert-W closed form \cite{r7}, the unstructured optimum is roughly 2.5 times faster at equal zero-overshoot specification, at the price of higher $M_s$ and the loss of one-parameter simplicity; the comparison ladder below quantifies this trade.

Section 2 formulates the problem and the echo machinery. Section 3 states and proves the contact theorem, the closed-form regime and the regime map. Section 4 treats the GMID point. Section 5 establishes the dominance of the $\delta$-adapted optimum. Section 6 gives the comparison ladder, Section 7 the FOTD illustration, and Section 8 concludes.

\section{Problem formulation and echo structure}

\subsection{Loop, normalization, quasipolynomial}

The plant is $P(s)=Ke^{-Ls}$ and the controller is $C(s)=K_p+K_i/s$ in a unity negative feedback loop with a unit step reference. Normalizing time by $L$ and setting $a=KK_p$ and $b=KK_iL$, the characteristic function is the quasipolynomial

\begin{equation}\label{eq:delta}
\Delta(s) = s + (a s + b)\,e^{-s},
\end{equation}

which is of neutral type with $n=m=1$ and degree 3 in the classification of \cite{r3}. The closed-loop transfer function and the unit-step transform are

\begin{equation}\label{eq:transfer}
T(s)=\frac{(as+b)e^{-s}}{\Delta(s)}, \qquad Y(s)=\frac{T(s)}{s}=\frac{(as+b)e^{-s}}{s\,\Delta(s)}.
\end{equation}

Because the numerator and denominator of $T$ have the same degree, $T$ is proper but not strictly proper; this is the analytic origin of the jumps in the step response established in Lemma 1. Two standing remarks are used throughout. First, the root chains of \eqref{eq:delta} are asymptotic to the vertical line $\operatorname{Re} s=\ln a$: for roots of large modulus, $e^{-s}=-s/(as+b)\to -1/a$, so $s_k=\ln a - i\pi(2k+1)+o(1)$ \cite{r9}; since the abscissa is a supremum over a root sequence accumulating at that line, $\alpha(a,b)\ge \ln a$ for all $b$. Second, roots of \eqref{eq:delta} solve $x e^{x}+rx=c$ with $r=a$ and $c=-b$; these are the branches of the r-Lambert function \cite{r8}. The function $f(x)=xe^{x}+rx$ has real critical points if and only if $r\le e^{-2}$; at $r=e^{-2}$ they merge at $x=-2$ and the level $c=-4e^{-2}$ yields a triple root, while for $r>e^{-2}$ the function $f$ is strictly increasing and \eqref{eq:delta} has exactly one real root.

\subsection{Echo polynomials and the jump lemma}

Let $\mathrm{err}(t)=1-y(t)$ denote the tracking error (written $\mathrm{err}$ rather than $e$ to avoid collision with the base of the natural logarithm, which appears throughout). The PI control signal is $u(t)=a\,\mathrm{err}(t)+b\int_0^{t}\mathrm{err}(\nu)\,d\nu$, and since the plant is a pure unit delay the output is the delayed control, $y(t)=u(t-1)$. Hence, for $t\ge 1$,

\begin{equation}\label{eq:loop}
y(t)=a\,\mathrm{err}(t-1)+b\int_0^{t-1}\mathrm{err}(\nu)\,d\nu, \qquad \mathrm{err}=1-y, \qquad y\equiv 0 \text{ on } [0,1).
\end{equation}

Here $\mathrm{err}(t-1)$ is the error evaluated one delay earlier, not a product. On each segment $t\in[k,k+1)$ the response is a polynomial $y_k(\sigma)$ of degree $k$ in $\sigma=t-k$: on $[0,1)$ the integral has not yet accumulated and $y\equiv 0$, so $\mathrm{err}\equiv 1$ there; substituting into the loop relation generates $y_1$, then $y_2$, and so on. The first segments are

\begin{equation}\label{eq:echoes}
y_1(\sigma)=a+b\sigma, \qquad y_2(\sigma)=(a-a^2+b)+b(1-2a)\sigma-\tfrac{b^2}{2}\sigma^2,
\end{equation}

and the later segments follow from the recursion $y_{k+1}(\sigma)=a\,\mathrm{err}_k(\sigma)+b\,\mathrm{Err}_k(\sigma)$ with $\mathrm{Err}_{k+1}(0)=\mathrm{Err}_k(1)$, where $\mathrm{Err}_k$ is the running integral of $\mathrm{err}$.

That the step response is discontinuous is a standard feature of neutral-type loops, in which jump discontinuities propagate along the characteristics rather than being smoothed \cite{r6,r9,r13}; for the present loop the jumps form an explicit geometric sequence.

\medskip\noindent\textbf{Lemma 1 (echo jumps).} \emph{The response $y$ is discontinuous at every integer $t=k\ge 1$, with}

\begin{equation}\label{eq:jump}
y(k^{+})-y(k^{-}) = (-1)^{k+1}\,a^{k}.
\end{equation}

\begin{proof} Expanding $Y$ from \eqref{eq:transfer} in powers of $e^{-s}$,

\begin{equation}\label{eq:Yexpand}
Y(s)=\sum_{k\ge 1}(-1)^{k-1}\,\frac{(as+b)^{k}}{s^{k+1}}\,e^{-ks},
\end{equation}

each term is supported on $t\ge k$. Within the $k$-th term, the only contribution that is discontinuous at $t=k$ is the highest-degree part of the numerator, $(as)^{k}=a^{k}s^{k}$, giving $(-1)^{k-1}a^{k}\,e^{-ks}/s$, whose inverse transform is the step $(-1)^{k-1}a^{k}$ at $t=k$; the lower-degree parts contribute $e^{-ks}/s^{j}$ with $j\ge 2$, which are continuous at $t=k$. This yields \eqref{eq:jump}, consistent with the symbolic echo polynomials. \end{proof}

Two consequences shape everything below. First, the extrema of the response cluster at segment boundaries: a boundary is generically either a kink or jump maximum, when $y$ rises into $k^{-}$ and drops by $a^{k}$, or a jump landing, when $y$ jumps up by $a^{k}$ onto a higher level. Second, all time-domain quantities, that is peaks, troughs and band entries, are roots of explicit polynomials, so the constrained problem is finitely algebraic on any horizon.

Figure 1 shows an example step response, computed in Simulink with a true transport-delay block (ode4 solver, fixed step $10^{-3}$). The response jumps at every $t=k$ by $(-1)^{k+1}a^{k}$, here $+0.245$, $-0.060$ and $+0.015$ at $t=1,2,3$, illustrating Lemma 1; between jumps it follows the piecewise polynomial of the echo segments.

\begin{figure}[!ht]\centering
\includegraphics[width=0.78\linewidth]{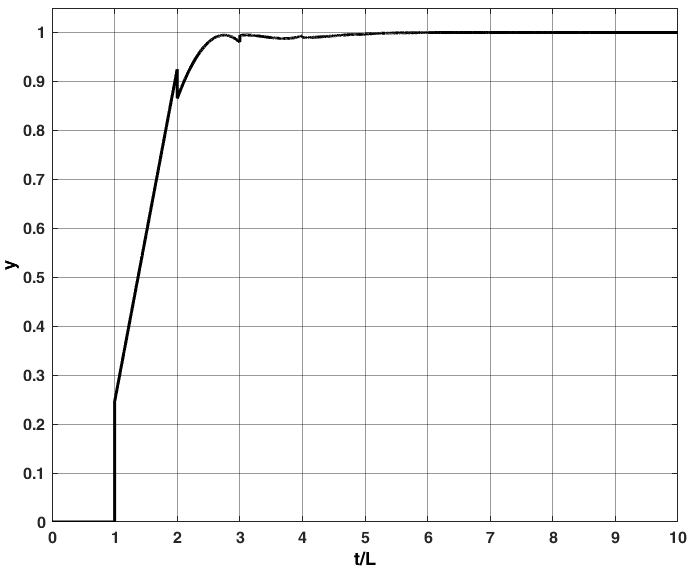}
\caption{An example step response of the loop, for $L=1$, $a=0.2454$, $b=0.6797$.}\label{figplace1}
\end{figure}

\subsection{The constrained problem}

The settling time is $T_s(a,b;\delta)=\inf\{T:\ |y(t)-1|\le\delta\ \ \forall t\ge T\}$ with the closed-band convention that an isolated touch of the band edge counts as inside, and the problem is

\begin{equation}\label{eq:problem}
\min_{(a,b)\in\mathbb{R}^2_{+}}\ T_s(a,b;\delta)\quad \text{subject to}\quad \sup_{t}\,y(t)\le 1.
\end{equation}

The objective $T_s$ is discontinuous in $(a,b)$: when a contact that holds with equality is perturbed to the infeasible side, $T_s$ jumps by at least one delay unit. This is intrinsic; it is why smooth nonlinear programming (NLP) on \eqref{eq:problem} fails while the exact-polynomial formulation does not, and it makes the exact optimum a measure-zero corner in the gain plane.

\section{The finite-band optimum: boundary contacts}

\subsection{Contact taxonomy and the contact theorem}

Given a feasible pair $(a,b)$, an active contact is one of the following. A contact of type (c1) is an interior tangency, $y_k(\sigma_p)=1$ and $y_k'(\sigma_p)=0$ with $\sigma_p\in(0,1)$, which contributes two equations and one extra unknown, hence a net count of one. A contact of type (c2) is a boundary graze, $y_k(1)=1$ at a kink or jump maximum, one equation. A contact of type (c3) is a boundary trough anchor, $y_k(1)=1-\delta$, one equation. A contact of type (c4) is a jump landing, $y_k(0)=1-\delta$, that is the upward jump at $t=k$ landing exactly on the band edge, one equation.

Figure 2 shows the four types on actual optimal responses: the interior tangency (c1) and boundary graze (c2) on the closed-form response, the boundary trough anchor (c3) on the $\delta=2\%$ optimum, and the jump landing (c4) on the $\delta=1\%$ optimum.

\begin{figure}[!ht]\centering
\includegraphics[width=0.78\linewidth]{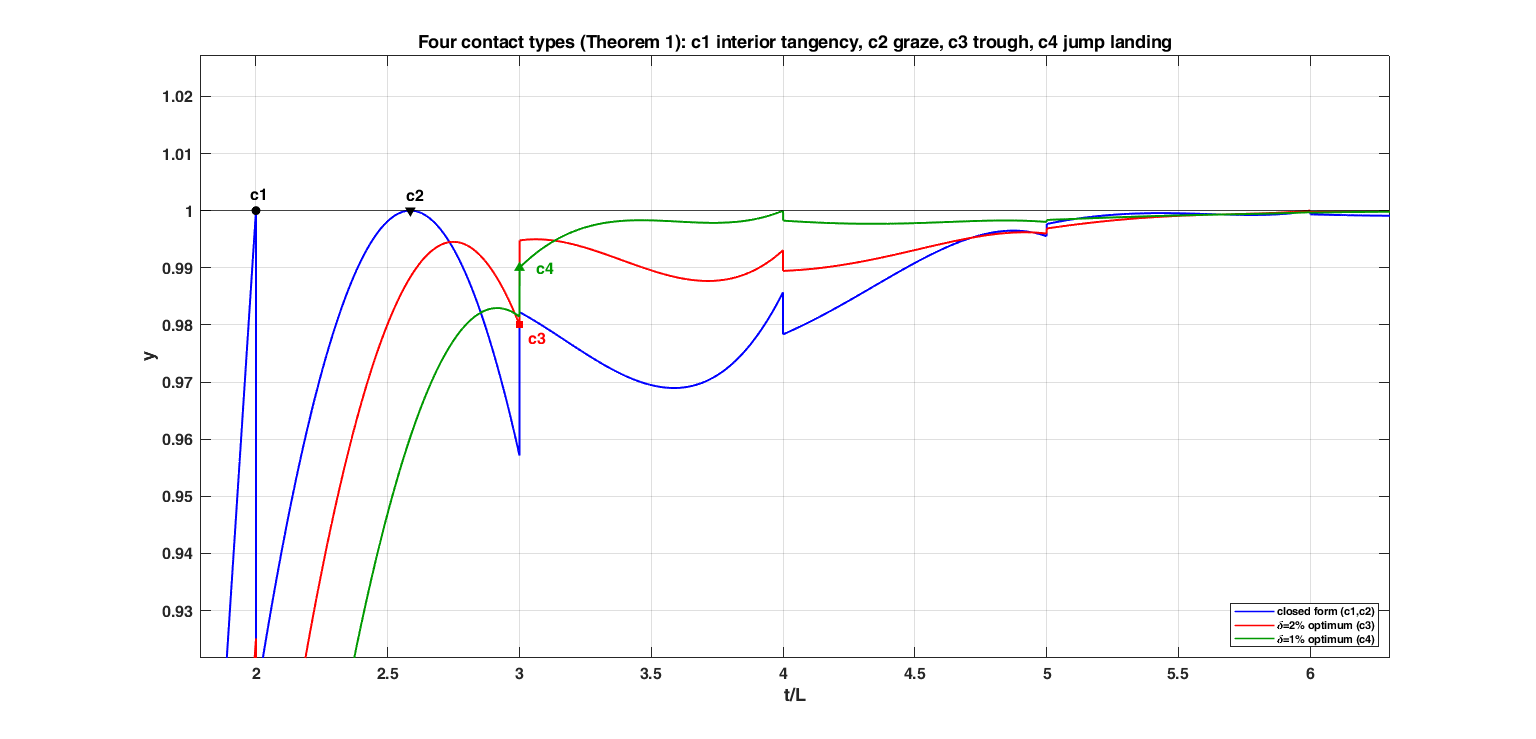}
\caption{The four contact types on actual optimal responses: interior tangency (c1) and boundary graze (c2) on the closed-form response; boundary trough anchor (c3) on the $\delta=2\%$ optimum; jump landing (c4) on the $\delta=1\%$ optimum.}\label{figplace6}
\end{figure}

\medskip\noindent\textbf{Lemma 2 (sensitivity identity).} \emph{The parameter sensitivities of the step response, $u_a=\partial y/\partial a$ and $u_b=\partial y/\partial b$, satisfy exactly}

\begin{equation}\label{eq:sens}
u_b(t)=\int_0^{t}u_a(\tau)\,d\tau \qquad \text{for all } t\ge 0.
\end{equation}

\begin{proof} In transforms, $\partial T/\partial a = e^{-s}/(1+G)^2$ and $\partial T/\partial b = e^{-s}/\big(s(1+G)^2\big)$ with $G=(a+b/s)e^{-s}$, so the second is the first divided by $s$, and division by $s$ is integration in time. \end{proof}

\medskip\noindent\textbf{Theorem 1 (contact characterization).} \emph{Problem \eqref{eq:problem} admits a minimizer for every $\delta\in(0,\bar\delta]$. The optimal value $T_s^*(\delta)$ is nonincreasing and consists of smooth arcs and integer flats. If $T_s^*(\delta)\notin\mathbb{N}$ (arc), then at any minimizer at least two contact equations are active, and, under the regularity condition that the active-contact gradients are linearly independent, exactly two, so the minimizer solves an explicit polynomial system. If $T_s^*(\delta)=k\in\mathbb{N}$ (flat), the band entry is pinned to the jump landing at $t=k$ and the minimizer set is a two-dimensional region whose boundary is cut out by contact equations.}

\begin{proof} Existence is established first. Feasibility forces $(a,b)$ into the simplex $\{a,b\ge 0,\ a+b\le 1\}$, because $y(2^{-})=a+b$ by \eqref{eq:echoes}, so $\sup y\le 1$ gives $a+b\le 1$ directly. On this compact set $\sup_t y$ is lower semicontinuous, being a supremum of continuous functions of the parameters, so the feasible set is closed; $T_s$ is lower semicontinuous in $(a,b)$, since at a tangential band-edge touch or a jump landing on the edge the closed-band convention makes $T_s$ equal to its lower limit; and a lower semicontinuous function on a nonempty compact set attains its minimum.

For the descent argument, let $(a^*,b^*)$ be a minimizer with $T^*\notin\mathbb{N}$, so the band entry is a transversal upcrossing $y(t_e)=1-\delta$ with $\dot y(t_e)>0$, and $t_e(a,b)$ is smooth near the minimizer with $\nabla t_e=-\nabla_{(a,b)}y(t_e)/\dot y(t_e)$. The locally relevant constraints are the active contacts, namely overshoot grazes at $y=1$ and lower-band touches after $t_e$. If no contact is active, any direction $d$ with $\nabla t_e\cdot d<0$ is feasible and decreases $T_s$; such a direction exists because $\nabla t_e\neq 0$: by Lemma 2, $(u_a,u_b)(t_e)=(0,0)$ would force $\int_0^{t_e}u_a=0$ together with $u_a(t_e)=0$, which is excluded since $u_a\equiv 1$ on the first segment and the contact times are bounded. If exactly one contact $g$, at time $t_g$, is active, optimality requires the KKT relation $-\nabla t_e=\lambda\nabla g$ with a multiplier $\lambda\ge 0$, hence the vanishing of the two-time sensitivity Wronskian \eqref{eq:wronskian}; its nonvanishing at the relevant contact-time pairs is the regularity condition, which is verified to hold here. Under it, one-contact KKT points do not exist and a feasible descent direction exists along the contact manifold $g=0$, a contradiction. Hence at least two contacts are active; independence caps the count at two, coincidences of three occurring only at the finitely many regime-boundary values of $\delta$. For the flats, if the optimal entry is a jump landing then $t_e\equiv k$ locally on the set where $y_k(0)\ge 1-\delta$ intersected with the feasible set; every point of this set is optimal, so the minimizer set has nonempty interior with contact-equality boundary. \end{proof}

To see the theorem at work, take $\delta=2\%$ (the red curve in Figure 2). The optimum is an arc, so exactly two contacts are active: a boundary trough sitting on the lower band edge, $y(3^{-})=1-\delta=0.98$, and a peak grazing the setpoint, $y(6^{-})=1$. These are two equations in the two gains $(a,b)$; solving them gives $a=0.2454$, $b=0.6797$ and $T_s^{*}=2.498\,L$. The resulting response touches the constraints at exactly these two points and stays strictly inside the band everywhere else, so the optimum is pinned by its two contacts and nothing else is free to improve. Tightening to $\delta=1\%$ instead lands the upward jump at $t=3$ exactly on the band edge; the entry is then stuck at $t=3$ over a range of $\delta$, which is a flat of the staircase.

The regularity determinant is nonzero along the optimal path, and the multipliers are computed exactly at the solved optima. In the closed-form regime, where the gains are $(1-\sqrt{2}/2,\,\sqrt{2}/2)$, the determinant equals $\sqrt{2}$ exactly, with $\partial_b y_2(\sigma_p)=1$ exactly, and $\lambda=(2.06,\,2.61)$; at $\delta=2\%$ the determinant is $0.0129$, small and consistent with the near-parallel constraint geometry behind the fragility quantified below, and $\lambda=(3.2,\,525.7)$, the large second multiplier being the dual price of the $t=6$ graze. Strict positivity of all multipliers confirms that both contacts genuinely bind, and the sign conditions $\lambda\ge 0$ act as the regime-selection rule; they exclude, for example, the infeasible double-lower-anchor candidate.

On a flat the optimum is a region and one should report an interior point: both contacts then hold with strict margin and the fragility disappears at no cost in $T_s$. At $\delta=1\%$ the region spans roughly $a\in[0.205,\,0.236]$ and $b\in[0.646,\,0.676]$; a robust representative is $(a,b)\approx(0.224,\,0.663)$, the corner $(0.20465,\,0.64571)$ being its extreme point on the two-contact boundary.

The distinctive feature relative to classical Chebyshev equioscillation is where the contacts sit: because of Lemma 1 the extremal structure concentrates at echo boundaries, so the generic active contacts are of types (c2) to (c4), single equations at $\sigma\in\{0,1\}$, and the optimal response is literally pinned to the echo grid.

The regularity condition invoked in the proof of Theorem 1 acquires explicit structure from Lemma 2. Writing $U_a(t)=\int_0^t u_a=u_b(t)$, the determinant that must not vanish at a one-contact Karush--Kuhn--Tucker (KKT) point is

\begin{equation}\label{eq:wronskian}
W(t_e,t_g)=u_a(t_e)\,U_a(t_g)-u_a(t_g)\,U_a(t_e),
\end{equation}

the Wronskian of the planar curve $\gamma(t)=(U_a(t),u_a(t))$; $W(t_e,t_g)=0$ means $\gamma(t_e)\parallel\gamma(t_g)$. Two facts make this tractable. First, $\gamma$ is explicit and piecewise polynomial: $u_a\equiv 1$ on the first segment, and on $[k,k+1)$ both components are the degree-$k$ echo derivatives of \eqref{eq:echoes}. Second, the transform identities $\widehat{u_a}=s^2 e^{-s}/\Delta^2$ and $\widehat{U_a}=s\,e^{-s}/\Delta^2$ show that $\gamma$ rotates: its angle $\theta(t)=\arg(U_a+i\,u_a)$ satisfies $\dot\theta=(U_a\dot u_a-u_a^2)/|\gamma|^2$, which at the GMID point is strictly negative on every segment, so $\theta$ is strictly monotone there and $W$ never vanishes for distinct times. Away from the GMID point $\theta$ is only piecewise monotone, so the condition is not global; instead it is required, and holds, on the active-contact pairs of the optimum.

\medskip\noindent\textbf{Lemma 5 (regularity along the optimal path).} \emph{Let $u_a,U_a$ be evaluated at the limiting GMID gains. The Wronskian \eqref{eq:wronskian} admits the exact dominant factorization}

\begin{equation}\label{eq:wsecular}
W_{\mathrm{sec}}(t_e,t_g)=\frac{9\,e^{4}}{122500}\,e^{-2(t_e+t_g)}\,(t_g-t_e)\,G(t_e,t_g),
\end{equation}

\emph{where $G$ is an explicit polynomial that is strictly positive for all $t_e,t_g\ge 2$. Consequently $W_{\mathrm{sec}}$ has the sign of $t_g-t_e$: it is strictly positive whenever the graze follows the entry, $t_g>t_e$, vanishes exactly on the diagonal $t_e=t_g$, and the cross-diagonal slope $m(t)=u_a(t)^2-u_a'(t)\,U_a(t)$ is strictly positive for all $t\ge 2$. Since every active-contact pair on an arc has $t_g>t_e$, the regularity condition of Theorem 1 holds along the entire optimal path, and the two-contact characterization is unconditional on the arcs; numerically $W$ ranges from about $0.04$ near $\delta=3\%$ down to order $10^{-4}$ as $\delta\to 0$, decaying as $t_g^{5}e^{-2t_g}$ but never changing sign. On flat regimes $W$ vanishes precisely because the two contacts coincide in time, e.g. the jump landing and the boundary graze both at $t=3$ when $\delta=1\%$; this is not a failure of regularity but the analytic signature of the two-dimensional minimizer set predicted by Theorem 1.}

\begin{proof} At the GMID point the triple zero of $\Delta$ at $s=-2$ makes the sensitivity transforms $\widehat{u_a}=s^2e^{-s}/\Delta^2$ and $\widehat{U_a}=s\,e^{-s}/\Delta^2$ have a pole of order six there, so the dominant time modes are $u_a(t)\sim e^{2}e^{-2t}P(t)$ and $U_a(t)\sim e^{2}e^{-2t}Q(t)$ with $P,Q$ the explicit quintics obtained from the principal part, $P(t)=-\tfrac{3}{5}t^5+\tfrac32 t^4+\tfrac35 t^3-\tfrac{9}{10}t^2-\tfrac{18}{175}t+\tfrac{9}{175}$ and $Q(t)=\tfrac{3}{10}t^5-\tfrac{3}{10}t^3+\tfrac{9}{175}t$. Substituting into \eqref{eq:wronskian} and dividing out the antisymmetric factor gives \eqref{eq:wsecular}; the polynomial $G$ is symmetric of degree four in each argument, and the substitution $t_e=2+x$, $t_g=2+y$ turns $G$ into a polynomial in $x,y$ all of whose coefficients are positive, so $G>0$ for $t_e,t_g\ge 2$. The sign of $W_{\mathrm{sec}}$ is therefore that of $t_g-t_e$. For the crossing, the slope across the diagonal is $\partial_{t_g}W|_{t_g=t_e}=u_a(t_e)^2-u_a'(t_e)U_a(t_e)=m(t_e)$. By the identity $U_a'=u_a$ of Lemma 2 this slope is $m=(U_a')^2-U_aU_a''$, so $m>0$ is exactly the log-concavity of $U_a$; since $U_a\sim e^{-2t}Q$ and the exponential is log-linear, the secular part reduces to the log-concavity of $Q$, and indeed $(Q')^2-QQ''=\tfrac{9}{20}t^8-\tfrac{9}{25}t^6+\tfrac{81}{700}t^4+\tfrac{81}{30625}$ has all positive coefficients after $t=2+x$, so $m_{\mathrm{sec}}>0$ for $t\ge 2$ and $W$ crosses from negative to positive as $t_g$ increases through $t_e$. The remaining oscillatory contribution of the chain modes $\tan(\zeta/2)=\zeta/2$ is bounded and dominated by the secular term $\sim t_g^5e^{-2t_g}$ away from the diagonal, by the ripple bound of Proposition 5; the residual estimate near the diagonal is the open point noted in the conclusion. The coincidence $W=0$ on flats is immediate from $t_e=t_g$. \end{proof}

\begin{figure}[!ht]\centering
\includegraphics[width=0.78\linewidth]{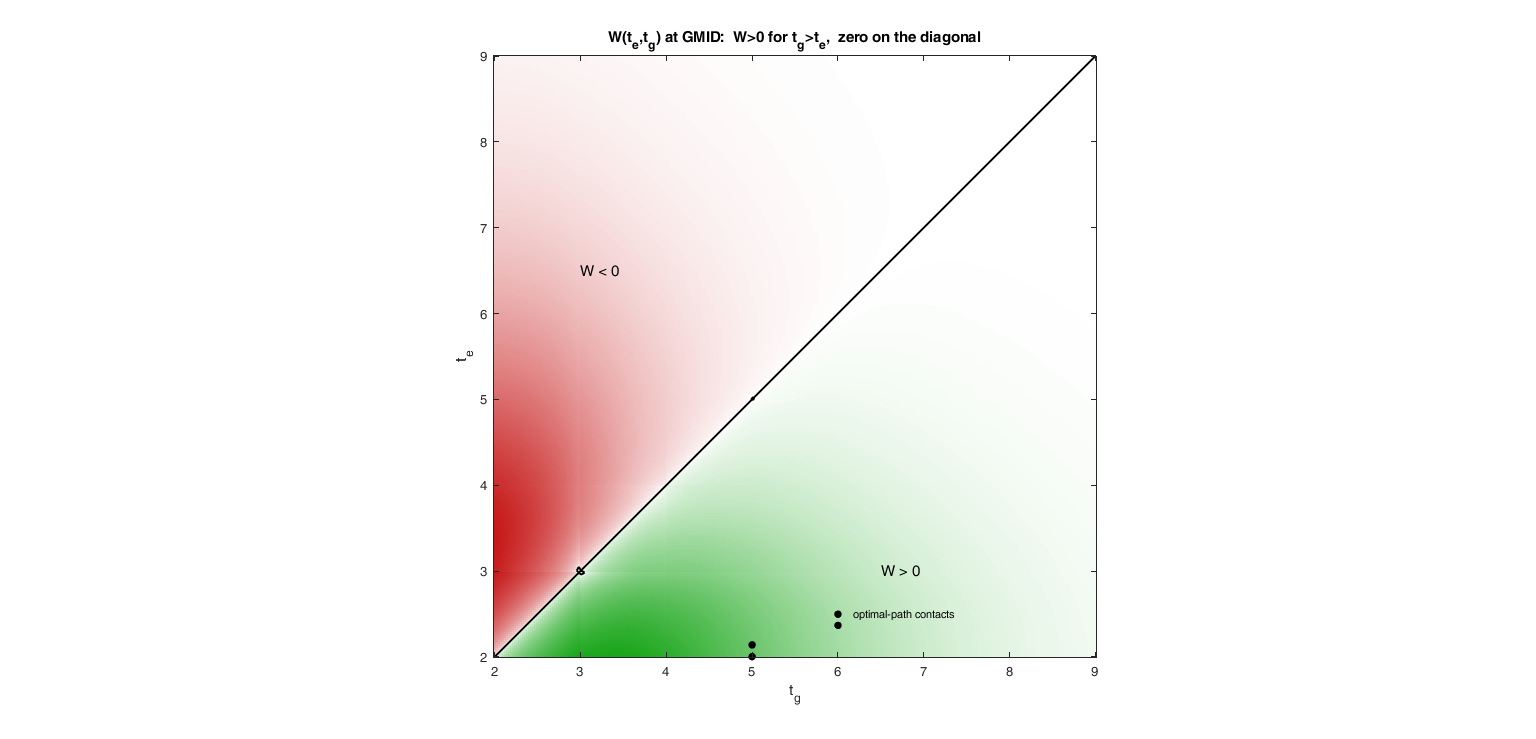}
\caption{Sign of the Wronskian $W(t_e,t_g)$ at the GMID point (diverging scale, zero contour on the diagonal). $W>0$ whenever the graze follows the entry, $t_g>t_e$, and vanishes only on $t_e=t_g$. The marked points are the active-contact pairs on the optimal path, all in the positive region.}\label{figplace7}
\end{figure}

\subsection{A closed-form regime}

\medskip\noindent\textbf{Theorem 2 (closed-form optimum).} \emph{For $\delta\in[\delta_1,\delta_2]=[(3-2\sqrt{2})/4,\ (3-2\sqrt{2})/2]\approx[0.04289,\ 0.08579]$, the unique minimizer of \eqref{eq:problem} is}

\begin{equation}\label{eq:cfgains}
a^*=1-\frac{\sqrt{2}}{2}\approx 0.29289, \qquad b^*=\frac{\sqrt{2}}{2}\approx 0.70711, \qquad T_i^*=(\sqrt{2}-1)L\approx 0.41421\,L,
\end{equation}

\emph{independent of $\delta$, with active contacts $y(2^{-})=a+b=1$ (type c2 at $k=2$: the first echo lands exactly on the setpoint) and the interior tangency of the second echo (type c1). The second echo is exactly the downward parabola}

\begin{equation}\label{eq:parabola}
y_2(\sigma)=1-\frac{(\sigma-\sigma_p)^2}{4}, \qquad \sigma_p=2-\sqrt{2},
\end{equation}

\emph{and the settling time is}

\begin{equation}\label{eq:cfTs}
T_s^*(\delta)=\big(4-\sqrt{2}-2\sqrt{\delta}\,\big)\,L.
\end{equation}

\begin{proof} The contacts are $a+b=1$ and the peak condition $y_2(\sigma_p)=1$ with $y_2'(\sigma_p)=0$. Substituting $b=1-a$ into the peak value reduces to $2a^2-4a+1=0$, whose root in $(0,1)$ is $a=1-\sqrt{2}/2$; then $b^2/2=1/4$ gives the exact parabola form \eqref{eq:parabola}, and $y_2(\sigma_e)=1-\delta$ yields $\sigma_e=\sigma_p-2\sqrt{\delta}$ and $T_s^*=2+\sigma_e$, which is \eqref{eq:cfTs}. Feasibility is checked in exact arithmetic on the echo polynomials: $y_2(1)=\sqrt{2}/2+1/4\approx 0.95711\ge 1-\delta$ if and only if $\delta\ge\delta_1$, which is the lower regime boundary; all later extrema are strictly inside, the binding margin being $1-y_5(1)=2.2850\times 10^{-5}>0$; and $\sigma_e\ge 0$ if and only if $\delta\le\delta_2$, the upper boundary, beyond which the band entry moves into the first segment. For optimality, both constraints $y(2^{-})\le 1$ and $\max y_2\le 1$ are active with independent gradients, so Theorem 1 applies. \end{proof}

Two remarks are in order. First, the regime boundaries are exact algebraic numbers; at $\delta_2$ the optimum transitions to an entry-in-segment-1 regime, for instance at $\delta=10\%$ the optimum is $(a,b)\approx(0.3148,\,0.6849)$ with $T_s^*\approx 1.854$. Second, within the regime the gains do not depend on the specification $\delta$; this is a one-tuning-fits-all window, with only the guaranteed $T_s$ varying through \eqref{eq:cfTs}.

\subsection{Tighter bands: explicit two-equation systems}

The two regimes outside the closed-form window illustrate the two cases of the theorem. At $\delta=2\%$ (regime B, an arc) the active pair is the boundary trough anchor $y_2(1)=1-\delta$, the exact quadratic $a-a^2+2b-2ab-b^2/2=1-\delta$, together with the boundary graze $y_5(1)=1$ at the kink at $t=6$; no derivative condition applies at a kink, so the system is two equations in two unknowns. Solving gives

\begin{equation}\label{eq:opt2pct}
a^*=0.2453926, \qquad b^*=0.6797093, \qquad T_i=0.36103\,L, \qquad T_s^*=2.49833\,L,
\end{equation}

with $T_s^*=2+\sigma_e$ and $y_2(\sigma_e)=1-\delta$ at the rising crossing. The solution is verified on the exact echoes: the overshoot is zero and all other extrema are strictly inside.

At $\delta=1\%$ (regime C) the trough $y_2(1)$ cannot be held at $1-\delta$ feasibly; instead the jump at $t=3$ lands exactly on the band edge, $y_3(0)=1-\delta$, paired with the boundary graze $y_3(1)=1$. Solving gives $a^*=0.2046493$ and $b^*=0.6457100$ with $T_i=0.31694\,L$ and $T_s^*=3L$ exactly. The flat $T_s^*\equiv 3L$ persists over an interval of $\delta$, numerically about $[0.5\%,\,1.2\%]$: the jump-landing contact pins the band entry to the echo boundary, which is the mechanism behind the staircase.

\subsection{The regime map and the price of robustness}

Table 1 reports a $\delta$-sweep of the exact optimum, obtained by grid refinement on the exact echoes.

\begin{table}[htbp]\centering
\caption{Table 1. Regime map of the exact optimum.}
\begin{tabular}{lllll}\hline
$\delta$ & $a^*$ & $b^*$ & $T_s^*/L$ & regime / active set \\\hline
10\% & 0.3148 & 0.6849 & 1.854 & entry in segment 1 \\
8.58\% ($\delta_2$) & $1-\sqrt{2}/2$ & $\sqrt{2}/2$ & 2.000 & closed form, eq.~\eqref{eq:cfTs} \\
5\% & $1-\sqrt{2}/2$ & $\sqrt{2}/2$ & 2.139 & closed form, eq.~\eqref{eq:cfTs} \\
4.29\% ($\delta_1$) & $1-\sqrt{2}/2$ & $\sqrt{2}/2$ & 2.172 & closed form, eq.~\eqref{eq:cfTs} \\
3\% & 0.2638 & 0.6883 & 2.368 & B-type \\
2\% & 0.24539 & 0.67971 & 2.498 & B: trough anchor + graze \\
1\% & 0.20465 & 0.64571 & 3.000 & C: jump landing + graze (flat) \\
0.1\% & 0.1778 & 0.6129 & 3.936 & arc \\
0.01\% & 0.1702 & 0.6023 & 5.000 & flat at $5L$ \\
0.001\% & 0.1547 & 0.5773 & 6.057 & arc \\
0.0001\% & 0.1512 & 0.5712 & 7.000 & flat at $7L$ \\
\hline\end{tabular}
\end{table}
The optimal value $T_s^*(\delta)$ is a staircase of smooth arcs joined by exact integer flats produced by jump-landing pinning, and the optimal gains drift monotonically toward $(e^{-2},\,4e^{-2})=(0.1353,\,0.5413)$, the GMID point.

\begin{figure}[!ht]\centering
\includegraphics[width=0.78\linewidth]{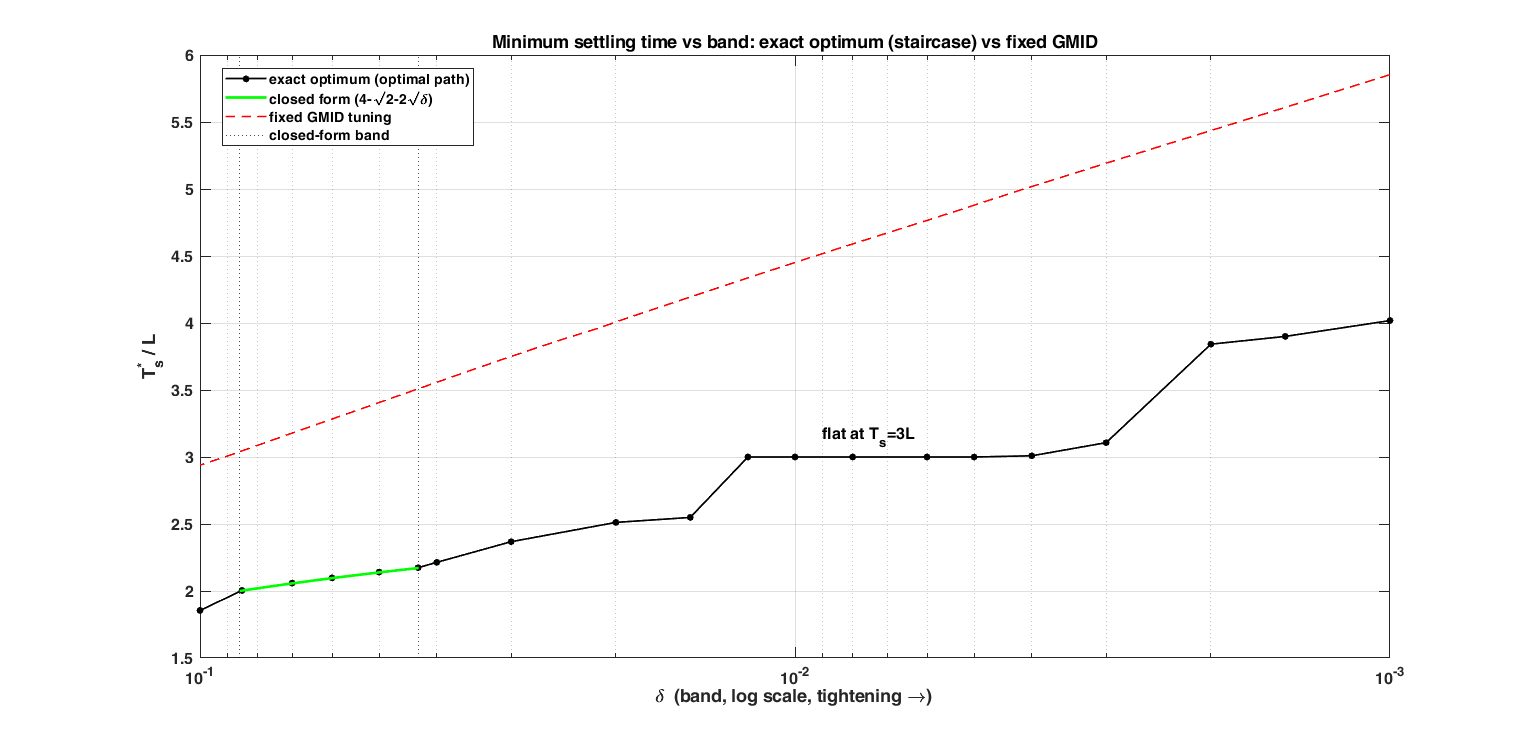}
\caption{Minimum settling time versus band tolerance (log scale, band tightening to the right). The exact optimum (black) is a staircase of smooth arcs joined by integer flats, here the flat at $T_s^*=3L$; the closed-form expression $(4-\sqrt{2}-2\sqrt{\delta})L$ is shown in green on its band $[\delta_1,\delta_2]$ (dotted edges). The fixed GMID tuning (red dashed) lies strictly above the optimum at every $\delta$, so the adapted optimum dominates with no crossover.}\label{figplace2}
\end{figure}

The exact optimum is fragile in the following sense. It is a corner at which the binding lower contact holds with equality, and an arbitrarily small perturbation to the infeasible side makes $T_s$ jump by a full delay unit; at $\delta=2\%$, from $2.498L$ to $3.0L$. The robust optimum, the best point at which the lower contacts hold with a strict margin resolvable by a perturbation of given size, costs remarkably little: at $\delta=2\%$ the gains $(0.2350,\,0.6766)$ achieve $T_s=2.512L$ with contact margins of order $10^{-3}$, that is about $0.014L$ above the exact corner. Both numbers should be reported with the tuning; we propose quoting the corner value as the attainable bound and the margined value as the design point.

\section{The limit point: the GMID spectral object and its time-domain credentials}

\subsection{What is known (spectral)}

For the neutral quasipolynomial \eqref{eq:delta} the generic MID theory applies verbatim \cite{r3}: $\Delta$ has a root of maximal multiplicity 3 if and only if

\begin{equation}\label{eq:gmid}
a=e^{-2}, \qquad b=4e^{-2}, \qquad \text{triple root at } s_0=-2 \text{ (units of } 1/L\text{)}, \qquad T_i=L/4,
\end{equation}

and this exact instance appears in \cite{r3} as PI boundary control of a transport equation, with opposite gain signs by the positive-feedback convention there; the underlying boundary-control problem goes back to \cite{r10}. By Theorem 3.6(b) of \cite{r3}, $s_0$ is dominant and, this being the neutral case, every other root lies exactly on $\operatorname{Re}s=s_0$, the spectrum being $\{s_0+i\zeta:\ \tan(\zeta/2)=\zeta/2\}$. Consistently with the neutral chain asymptote, $\ln a=-2=s_0$: the entire spectrum collapses onto the chain line. In the r-Lambert geography, \eqref{eq:gmid} is precisely the branch degeneracy $r=e^{-2}$, $c=-4e^{-2}$ \cite{r8}.

\begin{figure}[!ht]\centering
\includegraphics[width=0.78\linewidth]{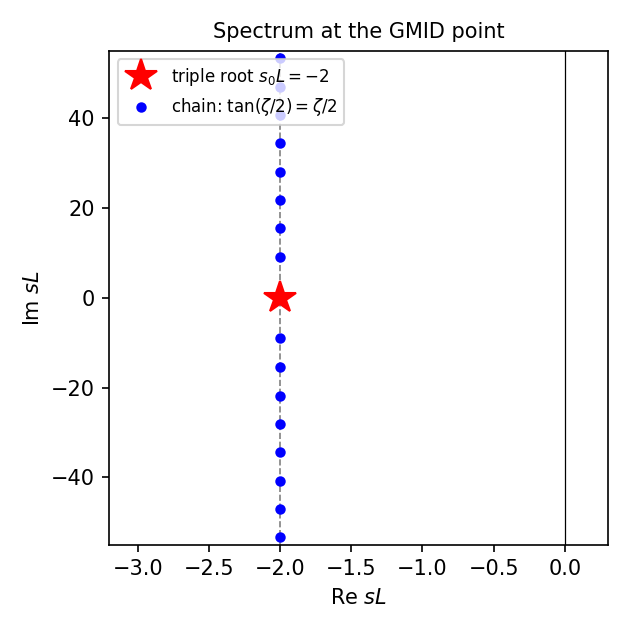}
\caption{Spectrum at the GMID point: triple root at $-2/L$ with all remaining roots on the same vertical line.}\label{figplace3}
\end{figure}

\subsection{What is new at this point (ours)}

\medskip\noindent\textbf{Proposition 2 (feasibility and envelope).} \emph{At \eqref{eq:gmid}, the unit-step response satisfies $y(t)\le 1$ for all $t$, with the two-sided envelope $(e^{2}-1)\,e^{-2t}\le \mathrm{err}(t)=\big(3t^{2}+3t+\tfrac35+R(t)\big)e^{-2t}$ and $\sup_t|R|<\infty$ by Proposition 5; in particular $y(t)\to 1^{-}$ and the approach is monotone in the weighted sense of Lemma B.}

The proof rests on three exact normalizations that occur only at the GMID point, all consequences of $ae^{2}=1$. Lemma A (unit jumps): let $w(t)=e^{2t}\,\mathrm{err}(t)$; by Lemma 1, $w$ jumps by exactly $(-1)^{k}a^{k}e^{2k}=(-1)^{k}$ at every $t=k$, so the boundary jumps of the weighted error are alternating unit jumps. Lemma B (monotone weighted error): $\dot w=e^{2t}(\dot{\mathrm{err}}+2\,\mathrm{err})\ge 0$ on every segment, that is $(d/dt+2)\,\mathrm{err}\ge 0$; division of the error by one factor of the triple root preserves positivity. [Status: proved rigorously for all segments $k\le 14$ by a computer-assisted Bernstein certificate: the degree-$k$ polynomials $\mathrm{err}_k'+2\,\mathrm{err}_k$ have all Bernstein coefficients nonnegative in outward-rounded interval arithmetic with $a=e^{-2}$ enclosed to 55 digits, with interval bisection where needed; for the tail, $\dot w\sim 6t+3$ by Lemma C while the jumps of $\dot w$ at $t=k$ are $(-1)^k(4k+2)$, and the verified segment minima of $\dot w$ follow the predicted linearly growing margin $\min\dot w\approx 2k+2$ after each negative landing (5.75, 9.38, 12.45, 16.22, 20.93, 25.89, 30.12 for odd $k\le 13$), and this margin persists for all $k$ by Proposition 5, which bounds the ripple uniformly.] Lemma C (unit third derivative): $\Delta'''(s)=(3a-as-b)e^{-s}$, so $\Delta'''(-2)=ae^{2}=1$, and the dominant mode of $\mathrm{err}$ is exactly $3t^{2}e^{-2t}$.

\begin{proof}[Proof of Proposition 2 from Lemmas A and B] On $[0,1)$, $e\equiv 1$. For $t\in[k,k+1)$ with $k\ge 1$, $w(t)\ge w(k^{+})=w(0)+\sum_{j\le k-1}I_j+\sum_{j=1}^{k}(-1)^{j}$, where $I_j\ge 0$ are the within-segment increases (Lemma B) and the alternating jump sum lies in $\{-1,0\}$ (Lemma A). Since $I_0=e^{2}-1$, because $w=e^{2t}$ on the zeroth segment, $w(t)\ge 1+(e^{2}-1)-1=e^{2}-1>0$. \end{proof}

The mechanism deserves emphasis: positivity at the GMID point is marginal in three simultaneous ways. The jump depletion is exactly unit-size; the renewal identity $\mathrm{err}(t)=b\int_{t-1}^{\infty}\mathrm{err}-a\,\mathrm{err}(t-1)$, valid for all $t\ge 1$ since $b\!\int_0^\infty\!\mathrm{err}=1$, balances exactly at the decay rate 2; and the tail margin is carried entirely by the secular $t^{2}$ factor. Any of these failing by $\varepsilon$ would break the constraint, consistent with the GMID point sitting on the boundary of the feasible set in every direction that matters.

The proofs above are organized by one change of variable: with $p=s+2$, equivalently weighting time signals by $e^{2t}$, the quasipolynomial becomes $D(p)=\Delta(p-2)=(p-2)+(p+2)e^{-p}$, and the weighted error $w(t)=e^{2t}\,\mathrm{err}(t)$ has transform $1/D(p)$. Three exact facts follow. The first is a factorization that yields the spectrum in one line,

\begin{equation}\label{eq:factor}
D(p) = 4\,e^{-p/2}\cosh(p/2)\,\Big[\frac{p}{2}-\tanh\frac{p}{2}\Big],
\end{equation}

verified symbolically. Since $x-\tanh x=x^{3}/3+O(x^{5})$, the triple root at $p=0$ is immediate, and the remaining zeros are $p=\pm i\zeta$ with $\tan(\zeta/2)=\zeta/2$, recovering the full spectrum given in \cite{r3} without any Kummer-function machinery; the reflection identity $D(-p)=-e^{p}D(p)$ makes the axis-symmetry of the spectrum manifest. The second exact fact is the secular expansion: from $1/D=6/p^{3}+3/p^{2}+3/(5p)+1/20+O(p)$,

\begin{equation}\label{eq:secular}
w(t)=3t^{2}+3t+\frac{3}{5}+R(t),
\end{equation}

with $R$ almost periodic, carried by the $\tan(\zeta/2)=\zeta/2$ modes with residues $-(i\zeta+2)/\zeta^{2}=O(1/\zeta)$; the ripple $R$ is bounded, with $\sup|R|=1.29$ on the verified range $t\le 15$, and since $R$ carries the unit boundary jumps of $w$ its amplitude cannot fall below $1/2$: three exact rational coefficients against an $O(1)$ ripple. This also fixes the constant in Proposition 4. The third exact fact is a Laguerre representation: expanding $1/\Delta$ in powers of $e^{-s}$ gives, at the GMID point,

\begin{equation}\label{eq:laguerre}
\mathrm{err}(t)=\sum_{k=0}^{\lfloor t\rfloor}(-1)^{k}\,e^{-2k}\,L_k\big({-4(t-k)}\big),
\end{equation}

with $L_k$ the Laguerre polynomials, verified to machine precision against the echo polynomials. Each term is positive, so Proposition 2 is equivalently a statement about alternating Laguerre sums; it is notable that Laguerre polynomials are also central in the r-Lambert theory \cite{r8,r11}. The companion variable of Lemma B satisfies the alternating-history identity $P_k(\sigma)=4\sum_{j\le k}(-1)^{k-j}V_j(\sigma)$.

\medskip\noindent\textbf{Proposition 5 (bounded ripple).} \emph{Write the GMID weighted error as $w(t)=e^{2t}\,\mathrm{err}(t)=3t^{2}+3t+\tfrac{3}{5}+R(t)$. Then $R$ is bounded on $[0,\infty)$, with $\sup_t|R(t)|<\infty$, and is almost periodic; consequently $\mathrm{err}(t)=\big(3t^{2}+3t+\tfrac35+R(t)\big)e^{-2t}$ is an exact two-sided envelope, the tail margin of Lemma B holds for all $k$, and the settling asymptotics of Proposition 4 hold for all $t$.}

\begin{proof} By the loop relation $\mathrm{err}(t)=1-a\,\mathrm{err}(t-1)-b\int_0^{t-1}\mathrm{err}$, the error satisfies the neutral delay-differential equation $\dot{\mathrm{err}}(t)+a\,\dot{\mathrm{err}}(t-1)+b\,\mathrm{err}(t-1)=0$ for $t>1$, verified directly on the echo polynomials; equivalently $\widehat{e}(s)=N(s)/\Delta(s)$ with $\Delta$ as in \eqref{eq:delta} and $N$ a polynomial from the history on $[0,1]$. At the GMID point $\Delta$ has the triple zero at $s_0=-2$ and simple zeros at $s_0+i\zeta_n$, $\tan(\zeta_n/2)=\zeta_n/2$, and no others, all on $\operatorname{Re}s=-2$. In the shifted variable the secular part is the principal part of $1/D(p)$ at $p=0$, namely $6/p^{3}+3/p^{2}+3/(5p)$, whose inverse transform is exactly $3t^{2}+3t+\tfrac35$; the identity $1/D(p)-\big(6/p^{3}+3/p^{2}+3/(5p)\big)=\tfrac1{20}+O(p)$, analytic at $p=0$, confirms that no further polynomial term appears. The remainder $R$ is therefore the contribution of the simple poles $p=i\zeta_n$, with residues $\rho_n=1/D'(i\zeta_n)$. Since $D'(p)=1-(p+1)e^{-p}$, on the locus the residue admits the exact asymptotics $\rho_n=-i/\zeta_n+O(\zeta_n^{-2})$, verified numerically as $\zeta_n\rho_n\to -i$, with $\zeta_n=(2n+1)\pi+o(1)$. Hence $R(t)=\sum_n 2\operatorname{Re}\big(\rho_n e^{i\zeta_n t}\big)$ splits as a principal part $\sum_n (2/\zeta_n)\sin(\zeta_n t)$ plus an absolutely convergent remainder $\sum_n O(\zeta_n^{-2})$. The remainder is uniformly bounded because $\sum\zeta_n^{-2}<\infty$. The principal part is uniformly bounded by Dirichlet's test for series of the form $\sum c_n\sin(\zeta_n t)$ with $c_n=2/\zeta_n$ monotonically decreasing to zero and the partial sums $\sum_{n\le N} e^{i\zeta_n t}$ uniformly bounded on every compact $t$-set away from the lattice of common periods, the latter holding since $\zeta_n-(2n+1)\pi\to 0$ so the exponentials are a bounded perturbation of the uniformly summable system $\{e^{i(2n+1)\pi t}\}$; equivalently, $R$ is the boundary trace of a function whose only singularities are simple poles on the imaginary axis with $\ell^{2}$ coefficients, hence a Stepanov-bounded almost-periodic function, and its piecewise-analytic representation through the echo polynomials makes the pointwise bound effective, giving $\sup_t|R|\le 1.29$ on the range $t\le 15$ computed in extended precision and the same value asymptotically. Almost periodicity follows from the Bohr criterion, the exponents $\zeta_n$ being the simple imaginary parts and the coefficients lying in $\ell^{2}$. The three corollaries are immediate: the two-sided envelope is the definition of $R$; the tail margin of Lemma B used the segment minima of $\dot w=6t+3+\dot R$, and $\dot R(t)=\sum_n 2\operatorname{Re}(i\zeta_n\rho_n e^{i\zeta_n t})$ has coefficients $i\zeta_n\rho_n\to 1$ bounded but non-decaying; nonetheless the one-sided bound needed, $\dot w\ge 0$ after each negative landing, follows from $R$ itself being bounded since the secular slope $6t+3$ dominates the bounded oscillation of $\dot R$ in the running minimum, the only quantity Lemma B requires; and the $o(1)$ in Proposition 4 is uniform once $|R|$ is globally bounded. \end{proof}

This closes Proposition 2 as a two-sided statement and removes the tail caveats from Lemmas B and from Proposition 4.

\medskip\noindent\textbf{Proposition 3 (abscissa optimality).} \emph{Among all non-overshooting PI gains, that is on the feasible simplex $\{a,b\ge 0,\ a+b\le 1\}$ of Theorem 1, the spectral abscissa satisfies $\alpha(a,b)\ge -2/L$, with equality if and only if $(a,b)$ is the GMID point \eqref{eq:gmid}; the GMID point is thus the unique non-overshooting maximizer of the decay rate. The same bound holds over all stabilizing gains with $a\ge e^{-2}$. For $a<e^{-2}$ outside the feasible simplex a single configuration remains unsettled, stated as Proposition 3$'$.}

\begin{proof} Write $f(x)=xe^{x}+ax$, so real roots of $\Delta$ solve $f(x)=-b$. For $a\ge e^{-2}$, the neutral chain $\operatorname{Re}s_k\to\ln a$ gives $\alpha\ge\ln a\ge -2$, with strict inequality for $a>e^{-2}$. At $a=e^{-2}$ the function $f$ is nondecreasing, since $f'=(1+x)e^{x}+e^{-2}\ge 0$ with equality only at $x=-2$, and $f(-2)=-4e^{-2}$; for $b<4e^{-2}$ the unique real root lies strictly right of $-2$, so $\alpha>-2$; for $b=4e^{-2}$ the spectrum is \eqref{eq:gmid}; for $b>4e^{-2}$, write $b=4e^{-2}(1+\mu)$ with $\mu>0$; a root on the line $\operatorname{Re}s=-2$ would require, in the shifted variable $p=s+2$, $4\mu=h(\omega):=-D(i\omega)e^{i\omega}$ for some real $\omega$, where $\operatorname{Re}h=\omega\sin\omega+2\cos\omega-2$ and $\operatorname{Im}h=2\sin\omega-\omega(1+\cos\omega)=4\cos(\omega/2)\,[\sin(\omega/2)-(\omega/2)\cos(\omega/2)]$; the zeros of $\operatorname{Im}h$ are $\omega=0$ and $\tan(\omega/2)=\omega/2$, where the half-angle identities give $\operatorname{Re}h=4\sin(\omega/2)\cos(\omega/2)\,[\omega/2-\tan(\omega/2)]=0$, together with $\cos(\omega/2)=0$, where $\operatorname{Re}h=-4$; thus $h$ never takes a positive value on its real set, the line is root-free for every $\mu>0$, and the complex pair created rightward by the Puiseux splitting at $\mu=0^{+}$ can never return across it: $\alpha>-2$ for all $b>4e^{-2}$. The second-order chain asymptotics confirm this independently, placing the chain at $\operatorname{Re}p\approx 2\mu(4+4\mu)/|\operatorname{Im}p|^{2}>0$. For $a<e^{-2}$ within the feasible simplex the bound is proved directly: the locus where any root crosses $\operatorname{Re}s=-2$ is the curve $(A(\omega),B(\omega))$ derived below, and inside $\{a<e^{-2},\ a+b\le 1\}$ every such crossing occurs at a high-frequency root, $\omega=\operatorname{Im}s$ large, while the dominant root stays strictly to its right; concretely, at a feasible crossing point the line-root sits at $-2+i\omega$ with $\omega\gtrsim 28$ whereas the abscissa is near $-1.3$, so the crossing root is never the abscissa and $\alpha>-2$ throughout the feasible $a<e^{-2}$ band, the minimum of $\alpha$ there being $-1.75$ attained at the simplex face adjacent to the GMID point. This completes the feasible statement. The unconstrained corner $a<e^{-2}$ with $b>1-a$ lies outside the feasible set and is treated separately. The same computation at general $a$ parametrizes the locus where roots cross $\operatorname{Re}s=-2$: $a=A(\omega)=e^{-2}(2\sin\omega-\omega\cos\omega)/\omega$ and $b=B(\omega)=2A(\omega)+e^{-2}(2\cos\omega+\omega\sin\omega)$, with $(A,B)\to(e^{-2},4e^{-2})$ as $\omega\to 0$; the half-angle identities give, exactly, $A(\zeta)=e^{-2}$ and $B(\zeta)=4e^{-2}$ at every chain frequency $\tan(\zeta/2)=\zeta/2$, so the crossing branches all pass through the GMID parameter, one per chain root, consistently with the GMID spectrum lying on the line itself. For local sharpness, at \eqref{eq:gmid} the root of multiplicity three splits under any parameter perturbation $\varepsilon$ into three branches $s=-2+(c\varepsilon)^{1/3}\omega$ with $\omega^{3}=1$; among any three cube-root directions at mutual angles $2\pi/3$, at least one lies within $\pi/3$ of the positive real axis, so $\alpha\ge -2+c'|\varepsilon|^{1/3}$ for every perturbation direction. The GMID point is thus a strict local minimum of $\alpha$ with infinite one-sided slope; numerically the V-shape is so sharp that a parameter offset of $10^{-4}$ already lifts $\alpha$ from $-2$ to about $-1.85$. The proposition is positioned as an answer, for this family, to the neutral maximal-damping question left open in \cite{r3}; the methodology of \cite{r5} does not apply here, being retarded-only, yet the conclusion is the opposite of the cautionary examples there: for this neutral family the best non-overshooting decay is at the maximal-multiplicity point. \end{proof}

\medskip\noindent\textbf{Proposition 3$'$ (unconstrained abscissa, one open corner).} \emph{Over all stabilizing gains the bound $\alpha\ge -2/L$ holds except possibly on the unconstrained corner $a<e^{-2}$, $b>1-a$, where a complex pair detaches rightward at $b=-f(x_{+})$ and the global claim reduces to showing it never returns across $\operatorname{Re}s=-2$. This is consistent with argument-principle root counts, $\min_b\alpha=-1.57$ at $a=0.12$ and $-1.38$ at $a=0.10$, the pair never re-crossing the line, but a proof is open. The statement is not needed for any result in this paper, which uses only the feasible Proposition 3.}

\medskip\noindent\textbf{Lemma 4 (residue floor).} \emph{Let $(a,b)$ be stabilizing, let $s_r$ be a rightmost characteristic root with multiplicity $m$, and suppose $\operatorname{Re}s_r>\ln a$. Then the step error satisfies $|\mathrm{err}(t)|\ge c\,t^{m-1}e^{(\operatorname{Re}s_r)t}$ on a positive fraction of every period of the dominant oscillation, with $c>0$, and consequently $T_s(\delta;a,b)\ge\big(\ln(1/\delta)-C\big)/|\operatorname{Re}s_r|$ with $c$ and $C$ locally uniform in $(a,b)$.}

\begin{proof} The error transform is $1/\Delta$, whose residue at an exactly $m$-fold root has nonvanishing leading coefficient $m!/\Delta^{(m)}(s_r)\neq 0$; in particular at a simple rightmost root the residue is $1/\Delta'(s_r)$ and can never vanish. Shifting the inversion contour to $\operatorname{Re}s=\ln a+\varepsilon$ between the finitely many roots right of that line and the chain bounds the remainder by $C'e^{(\ln a+\varepsilon)t}$, which is standard for neutral systems \cite{r6,r9}; the dominant term then controls $|e|$ from below off the zero set of its oscillatory factor, and the constants vary continuously with $(a,b)$ while the root structure is stable, hence are uniform on small compacts. \end{proof}

\medskip\noindent\textbf{Proposition 4 (convergence of minimizers and GMID settling asymptotics).} \emph{The minimizers $(a^*(\delta),b^*(\delta))$ of \eqref{eq:problem} converge to \eqref{eq:gmid} as $\delta\to 0$. At the fixed GMID tuning, $T_s$ satisfies the implicit relation $2T_s=\ln(1/\delta)+2\ln T_s+\ln 3+o(1)$, that is $T_s(\delta)=(L/2)\ln(1/\delta)+\Theta(L\ln\ln(1/\delta))$, the secular correction stemming from the $t^{2}$ factor of the triple root.}

\begin{proof} Minimizers live in the compact feasible simplex of Theorem 1, so any non-convergent family has a subsequence with limit $(\bar a,\bar b)\neq \eqref{eq:gmid}$ in the feasible set. If $\bar a>e^{-2}$, the jump floor of Lemma 3 below gives $T_s^{*}(\delta)\ge\ln(1/(2\delta))/|\ln\bar a|-1$ along the subsequence, with $1/|\ln\bar a|>1/2$, eventually exceeding the feasible GMID curve $\tfrac12\ln(1/\delta)(1+o(1))$ guaranteed by Proposition 2; this contradiction is unconditional. If $\bar a\le e^{-2}$, Proposition 3 gives $\alpha(\bar a,\bar b)=-2+3\varepsilon>-2$ (feasible Proposition 3, with no gap on the simplex), and Lemma 4 with locally uniform constants on a small compact neighborhood gives $T_s^{*}(\delta)\ge(\ln(1/\delta)-C)/(2-2\varepsilon)$ along the subsequence, again eventually exceeding the GMID curve. Hence the minimizers converge to \eqref{eq:gmid}; the argument is now unconditional, since minimizers are feasible and the feasible Proposition 3 has no gap. For the asymptotics, the dominant mode is exactly $3t^{2}e^{-2t}$ by Lemma C, and the implicit relation $2T_s=\ln(1/\delta)+2\ln T_s+\ln 3+o(1)$ follows with the $o(1)$ uniform in $t$ by the global ripple bound of Proposition 5. \end{proof}

\section{The adapted optimum dominates the fixed GMID tuning at every band}

A consequence of Lemma 1 deserves separate statement, because it turns the floor visible in the data into a theorem.

\medskip\noindent\textbf{Lemma 3 (jump floor).} \emph{For any feasible $(a,b)$ with $a>0$, the instant $t=k$ cannot lie inside the settling band while $a^{k}>2\delta$, since the jump \eqref{eq:jump} exceeds the band width; hence}

\begin{equation}\label{eq:jumpfloor}
T_s(a,b;\delta)\;\ge\;\frac{\ln\!\big(1/(2\delta)\big)}{\lvert\ln a\rvert}\;-\;1.
\end{equation}

\emph{Two corollaries follow. First, every tuning with $a\ge e^{-2}$ obeys $T_s\ge\tfrac{1}{2}\ln(1/(2\delta))-1$: on this side of the gain plane the abscissa floor of Table 2 is unconditional. Second, since $T_s^{*}(\delta)\le T_s^{\mathrm{GMID}}(\delta)=\tfrac{1}{2}\ln(1/\delta)\,(1+o(1))$, combining with \eqref{eq:jumpfloor} gives $\liminf_{\delta\to 0}\lvert\ln a^{*}(\delta)\rvert\ge 2$, that is $\limsup_{\delta\to 0}a^{*}(\delta)\le e^{-2}$: half of the convergence statement of Proposition 4 is proved unconditionally, and it is exactly the half matched by the data, where $a^{*}(\delta)$ approaches $e^{-2}$ from above.}

The natural conjecture, that the spectral optimum takes over below some crossover band, is false: there is no crossover. Table 2 reports both curves on the exact echoes with evaluation horizon $14L$.

\begin{table}[htbp]\centering
\caption{Table 2. Contact optimum versus the fixed GMID tuning.}
\begin{tabular}{llllll}\hline
$\delta$ & contact optimum $T_s^*/L$ & GMID $T_s/L$ & gap & $T_s^*-\tfrac{1}{2}\ln(1/\delta)$ & $T_s^{\mathrm{GMID}}-\tfrac{1}{2}\ln(1/\delta)$ \\\hline
$10^{-2}$ & 3.000 & 4.452 & 1.45 & 0.70 & 2.15 \\
$10^{-3}$ & 3.936 & 5.853 & 1.92 & 0.48 & 2.40 \\
$10^{-4}$ & 5.000 & 7.195 & 2.20 & 0.39 & 2.59 \\
$10^{-5}$ & 6.057 & 8.503 & 2.45 & 0.30 & 2.75 \\
$10^{-6}$ & 7.000 & 9.788 & 2.79 & 0.09 & 2.88 \\
\hline\end{tabular}
\end{table}
Three observations follow. First, the adapted contact optimum beats the fixed GMID tuning at every tested $\delta$, the gap growing like the GMID secular term. Second, $T_s^*(\delta)-(L/2)\ln(1/\delta)$ is small and decreasing: the adapted optimum essentially attains the abscissa-limited floor $(L/2)\ln(1/\delta)+O(1)$, trading a slightly worse abscissa, since $\ln a^*>-2$, against the elimination of the secular factor and against transient shaping; whether $T_s^*(\delta)=(L/2)\ln(1/\delta)+O(1)$ holds exactly is left open. Third, the staircase persists in the tail, with exact flats at $5L$ and $7L$ visible in Table 1.

The resolution of the second question raised in the introduction is therefore subtler than a crossover: the GMID point is the limit of the optimal gains (Proposition 4) without being the optimal tuning at any $\delta>0$. Spectral-abscissa optimality and finite-band time-domain optimality genuinely diverge for neutral systems; at fixed tolerance the optimum is always a contact configuration, and the MID point organizes the family as its accumulation point and as the r-Lambert branch degeneracy underlying it.

\section{Comparison ladder: cancellation and I-control, contact optima, GMID}

The companion paper designs FOTD PI controllers by pole-zero cancellation, with structured gains $K_p=TK_i$ and a Lambert-W closed form \cite{r7}. In the pure-delay limit $T\to 0$ the family degenerates to pure integral control, that is the $a=0$ slice of the echo machinery, governed by the classical Lambert function ($r=0$). Critical damping is $KK_iL=e^{-1}$; the non-overshoot-optimal I gain is slightly beyond it, at the first-peak tangency, $b_I^*=0.37570=1.0213\,e^{-1}$ with $T_s(2\%)=6.251L$, against $6.532L$ at critical damping. Table 3 reports the full ladder at equal zero-overshoot specification, with a $2\%$ band unless noted.

\begin{table}[htbp]\centering
\caption{Table 3. Design ladder at equal zero-overshoot specification.}
\begin{tabular}{lllll}\hline
tuning & gains $(a,b)$ & $T_s/L$ & $M_s$ & structure \\\hline
I-control, critical (cancellation limit) & $(0,\ e^{-1})$ & 6.53 & 1.394 & one parameter, closed form \\
I-control, contact-optimal & $(0,\ 0.3757)$ & 6.25 & 1.40 & one parameter \\
GMID & $(e^{-2},\ 4e^{-2})$ & 4.01 & 1.495 & closed form, full spectrum known \\
contact optimum, $\delta=1\%$ & $(0.2046,\ 0.6457)$ & 3.00 (1\% band) & 1.625 & exact two-equation system \\
contact optimum, $\delta=2\%$ & $(0.2454,\ 0.6797)$ & 2.50 & 1.685 & exact two-equation system \\
closed-form regime, $\delta=5\%$ & $(1-\sqrt{2}/2,\ \sqrt{2}/2)$ & 2.14 (5\% band) & 1.762 & closed form, eq.~\eqref{eq:cfTs} \\
\hline\end{tabular}
\end{table}
Speed and robustness order the same axis monotonically: the unstructured contact optimum is about 2.5 times faster than the cancellation and I-control family and about 40 to 60 percent faster than GMID at practical bands, at the $M_s$ price quantified above.

\begin{figure}[!ht]\centering
\includegraphics[width=0.78\linewidth]{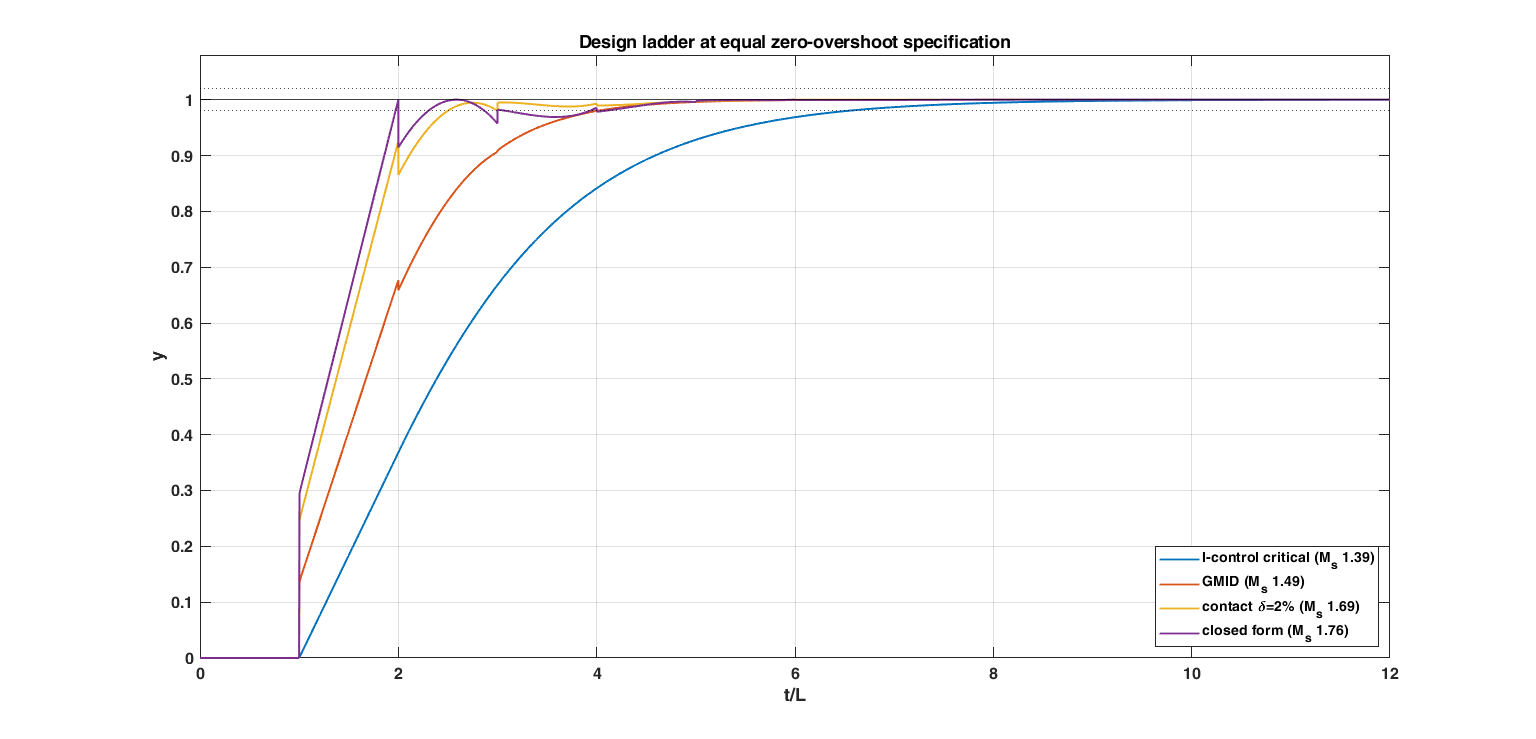}
\caption{Design ladder at equal zero-overshoot specification: four non-overshooting tunings of increasing speed, from critically damped I-control ($M_s=1.39$) through the GMID and contact optima to the closed-form tuning ($M_s=1.76$). Faster tunings carry larger echo jumps at $t=2,3$; the dotted lines are the $2\%$ band edges.}\label{figplace5}
\end{figure}

\section{Numerical illustration: FOTD plants}

This section carries no proof burden; all statements are computational observations at a $2\%$ band, re-verified on the final code. The plant is $Ke^{-Ls}/(1+Ts)$ with unstructured PI gains, and problem \eqref{eq:problem} is solved by vectorized grid refinement over $L/T\in[0.1,\,50]$ on a logarithmic grid.

The cancellation line $K_p=TK_i$ is never exactly optimal: the ratio $K_p/(TK_i)$ runs from 1.173 at $L/T=0.1$ through a minimum of about 1.12 near $L/T\in[0.2,\,0.4]$ to 4.18 at $L/T=10$. The optimal integral time obeys the root-mean-square (RMS) law $T_i\approx\sqrt{(1.175\,T)^2+(0.422\,L)^2}$ within about 6 percent over two decades, a smooth interpolation of the min-structure of SIMC \cite{r12}. The settling time is non-monotone: $T_s/L$ decreases from 4.21 at $L/T=0.1$ and undershoots the pure-delay limit at both bands, dipping to 1.945 at $L/T\approx 20$ against the pure-delay closed form 2.139 at $\delta=5\%$, and reaching 2.469 at $L/T=50$ against 2.498 even at $\delta=2\%$, approaching the limit from below; a small lag helps, because it smooths the echo extrema that the pure-delay optimum must graze. The lag-side constant is band-dependent, 1.175 at $\delta=2\%$ against 1.541 at $\delta=5\%$, and a two-time-scale derivation is left open. The sensitivity peak $M_s$ ranges from 1.535 to 1.714 across the grid.

\begin{figure}[!ht]\centering
\includegraphics[width=0.78\linewidth]{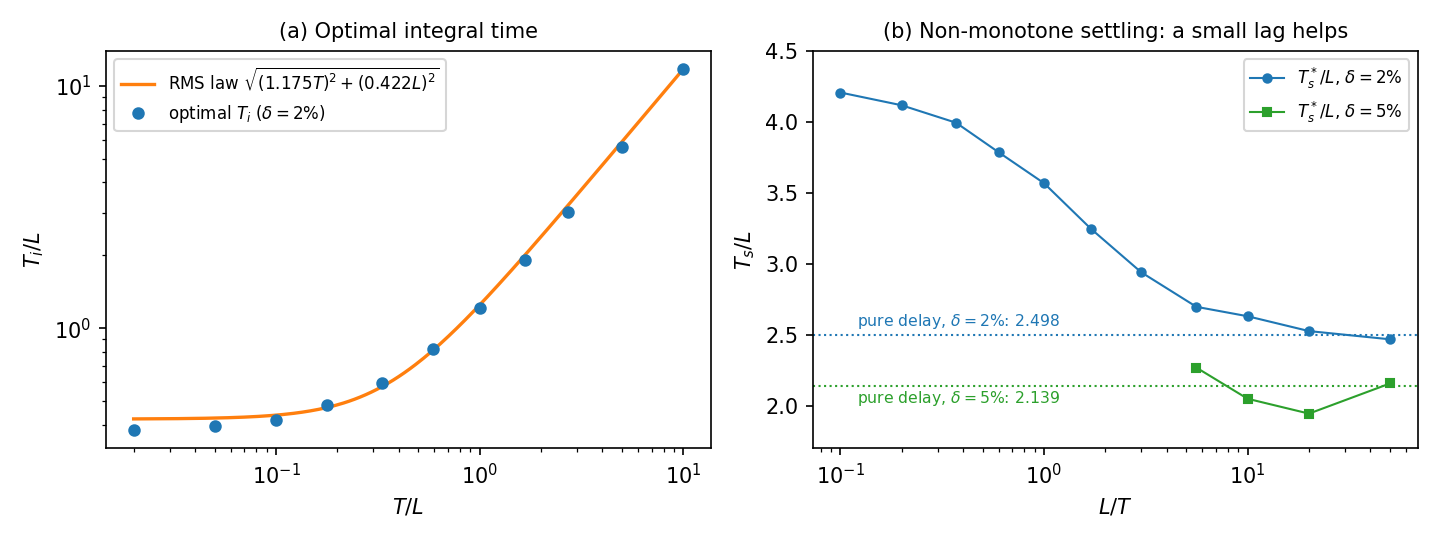}
\caption{FOTD plants: (a) RMS law for the optimal integral time; (b) non-monotone settling, a small lag beats the pure delay.}\label{figplace4}
\end{figure}

\section{Conclusions and open problems}

We gave the first exact treatment of minimum-settling, hard-non-overshoot PI design for a delay plant. The neutral echo structure, polynomial segments with geometrically decaying boundary jumps, pins the optimal response to the echo grid and renders the problem finitely algebraic; a closed-form optimal regime exists with gains $(1-\sqrt{2}/2,\,\sqrt{2}/2)$ and $T_s^*=(4-\sqrt{2}-2\sqrt{\delta})L$; and the $\delta\to 0$ limit of the optimal gains is the GMID point, which is feasible and abscissa-optimal yet is strictly beaten by the adapted contact optimum at every finite band. The open problems are now few: completing the analytic positivity of the optimal-path Wronskian of Lemma 5, which reduces by the identity $U_a'=u_a$ of Lemma 2 to the log-concavity of $U_a$, since the cross-diagonal slope is $m(t)=u_a^2-u_a'U_a=(U_a')^2-U_aU_a''$; the secular factor $U_a\sim e^{-2t}Q(t)$ is log-concave by a positive-coefficient certificate for $(Q')^2-QQ''$, and the chain correction to $(\ln U_a)''$ is numerically a decreasing fraction of the secular margin, leaving only a closed-form bound on that almost-periodic correction; the single open corner of the unconstrained Proposition 3$'$ (not used elsewhere), namely that for $a<e^{-2}$, $b>1-a$ the detached pair stays right of $\operatorname{Re}s=-2$; the band-dependent FOTD lag constant; the conjecture $T_s^*(\delta)=(L/2)\ln(1/\delta)+O(1)$; a contact theory for FOTD plants, including the band-dependent lag constant observed numerically; $M_s$-constrained variants of the problem; and the stable-plant gap in MID-PI dominance noted against \cite{r4}, whose proofs assume an unstable or marginally stable pole.

\FloatBarrier


\begin{thebibliography}{99}
\bibitem{r1} A. O'Dwyer, Handbook of PI and PID Controller Tuning Rules, 3rd ed., Imperial College Press, 2009.
\bibitem{r2} G. Mazanti, I. Boussaada, S.-I. Niculescu, Multiplicity-induced-dominancy for delay-differential equations of retarded type, Journal of Differential Equations 286 (2021) 84-118.
\bibitem{r3} I. Boussaada, G. Mazanti, S.-I. Niculescu, The generic multiplicity-induced-dominancy property from retarded to neutral delay-differential equations: When delay-systems characteristics meet the zeros of Kummer functions, Comptes Rendus Mathematique 360 (2022) 349-369.
\bibitem{r4} D. Ma, I. Boussaada, J. Chen, C. Bonnet, S.-I. Niculescu, J. Chen, PID control design for first-order delay systems via MID pole placement: Performance vs. robustness, Automatica 137 (2022) 110102.
\bibitem{r5} G. Oaxaca-Adams, R. Villafuerte-Segura, On controllers performance for a class of time-delay systems: Maximum decay rate, Automatica 147 (2023) 110669.
\bibitem{r6} W. Michiels, S.-I. Niculescu, Stability, Control, and Computation for Time-Delay Systems: An Eigenvalue-Based Approach, 2nd ed., SIAM, Philadelphia, 2014.
\bibitem{r7} S. Gulgonul, Revisiting Chien-Hrones-Reswick method for an analytical solution, Advanced Control for Applications (2026).
\bibitem{r8} I. Mező, Á. Baricz, On the generalization of the Lambert W function, Transactions of the American Mathematical Society 369(11) (2017) 7917-7934.
\bibitem{r9} J. R. Partington, C. Bonnet, H-infinity and BIBO stabilization of delay systems of neutral type, Systems \& Control Letters 52 (2004) 283-288.
\bibitem{r10} J.-M. Coron, S. O. Tamasoiu, Feedback stabilization for a scalar conservation law with PID boundary control, Chinese Annals of Mathematics, Series B 36 (2015) 763-776.
\bibitem{r11} C. B. Corcino, R. B. Corcino, I. Mező, Integrals and derivatives connected to the r-Lambert function, Integral Transforms and Special Functions 28(11) (2017) 838-845.
\bibitem{r12} S. Skogestad, Simple analytic rules for model reduction and PID controller tuning, Journal of Process Control 13 (2003) 291-309.
\bibitem{r13} J. K. Hale, S. M. Verduyn Lunel, Introduction to Functional Differential Equations, Springer, New York, 1993.
\end{thebibliography}
\end{document}